\documentclass[12pt]{extarticle}

\usepackage{preamble/custom_diagrams}
\usepackage{preamble/miscellaneous}
\usepackage{preamble/definitions}
\usepackage{preamble/text_formatting}
\usepackage{preamble/bibliography}

\title{\vspace{-1.0cm}\MetaTitle}
\date{\today}
\author{\MetaAuthor}

\begin{document}

\maketitle

\begin{abstract}
  Neural networks are now deployed in a wide number of areas from object classification to natural language systems.
  Implementations using analog devices like memristors promise better power efficiency, potentially bringing these applications to a greater number of environments.
  However, such systems suffer from more frequent device faults and overall, their exposure to adversarial attacks has not been studied extensively.
  In this work, we investigate how nonideality-aware training---a common technique to deal with physical nonidealities---affects adversarial robustness.
  We find that adversarial robustness is significantly improved, even with limited knowledge of what nonidealities will be encountered during test time.
\end{abstract}

\section{Introduction}

Modern machine learning has enabled a new era of software development that goes beyond rules-based programming.
Structures like neural networks can now tackle less well-defined, cognitive tasks in image recognition, natural language processing, and other areas.
Such systems have contributed to the development of applications ranging from voice recognition~\cite{radfordRobustSpeechRecognition} to autonomous driving~\cite{grigorescuSurveyDeepLearning2020} to chatbots~\cite{openaiGPT4TechnicalReport2024}.

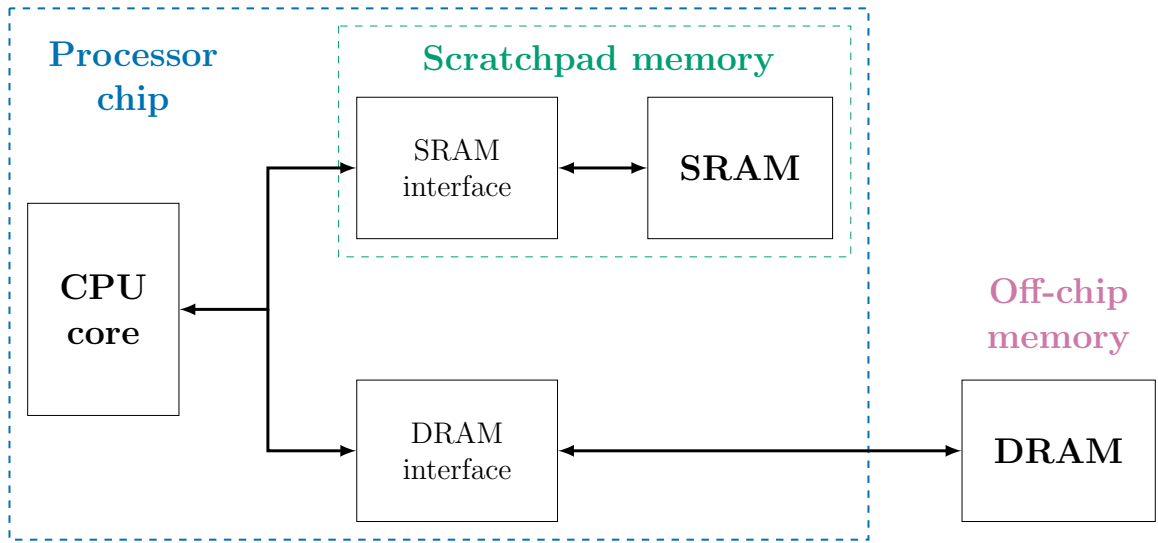
\begin{figure}[h]
    \centering
  \tikzsetnextfilename{tikz--sram-dram}%
  \begin{tikzpicture}
  \tikzstyle{unit}=[draw=black, align=center, minimum height=2cm]
  \tikzstyle{main-unit}=[unit, font=\large\bfseries, inner xsep=0.45cm]
  \tikzstyle{interface-unit}=[unit, inner xsep=0.65cm]
  \tikzstyle{data-flow}=[very thick]

  \node[main-unit, minimum height=3cm] (cpu-core) at (0, 0) {CPU\\core};
  \node[interface-unit] (sram-interface) at (5, 2) {SRAM\\interface};
  \node[main-unit, right of=sram-interface, xshift=3cm] (sram) {SRAM};
  \node[interface-unit] (dram-interface) at (5, -2) {DRAM\\interface};
  \node[main-unit, right of=dram-interface, xshift=7.5cm] (dram) {DRAM};

  \draw[dashed, draw=okabe-green] ($(sram-interface.north west) + (-0.25, 1)$) rectangle ($(sram.south east) + (0.25, -0.25)$);
  \draw[thick, dashed, draw=okabe-blue] ($(cpu-core.west |- sram-interface.north) + (-0.25, 1.25)$) rectangle ($(sram.east |- dram.south) + (0.5, -0.25)$);

  \node[font=\large\bfseries, text=okabe-green] at ($0.5*(sram-interface.north) + 0.5*(sram.north) + (0, 0.5)$) {Scratchpad memory};
  \node[font=\large\bfseries, align=center, text=okabe-blue] at ($(cpu-core.west |- sram-interface.north) + (1.5, 0.25)$) {Processor\\chip};
  \node[above of=dram, yshift=0.9cm, font=\large\bfseries, align=center, text=okabe-reddish-purple] {Off-chip\\memory};

  \draw[data-flow, latex-latex] (cpu-core.east) -- ++(1.25, 0) coordinate(intersection) -- (\currentcoordinate |- sram-interface.west) -- (sram-interface.west);
  \draw[data-flow, latex-latex] (sram-interface) -- (sram);
  \draw[data-flow, -latex] (intersection) -- (\currentcoordinate |- dram-interface.west) -- (dram-interface.west);
  \draw[data-flow, latex-latex] (dram-interface) -- (dram);
\end{tikzpicture}%

    \caption{
      Von~Neumann architecture.
      The arrows denote the data flow.
      Adapted from Ref.~\cite{mehonicEmergingNonvolatileMemories2023}.
    }
    \label{fig:sram-dram}
\end{figure}

However, one of the biggest issues, is the time and energy costs related to both the training and inference stages of these statistical models.
The main cause of this is the von~Neumann bottleneck---the separation of memory and processing units in traditional computers, shown in Figure~\ref{fig:sram-dram}.
Large models may need to be stored and accessed from off-chip memory (like DRAM), which incurs significant latency and energy costs.
Even if the model fits on on-chip memory (like SRAM), the data still needs to be repeatedly moved, causing inefficiency.
High costs stemming from these inefficiencies constrain who can afford to train and run these models, and limit the environments in which they can be deployed.
This is especially relevant due to the rapid penetration of machine learning at the edge, e.g., the Internet of Things, where there exists a need for fast and efficient inference on less-capable hardware.

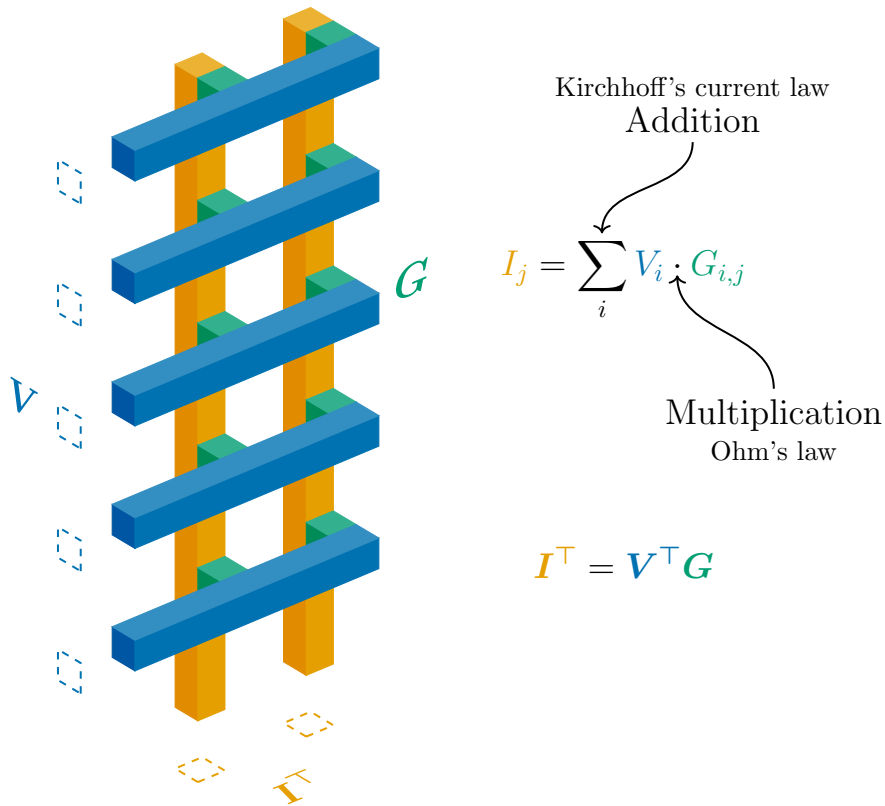
\begin{figure}[h]
    \centering
  \tikzsetnextfilename{tikz--crossbar-array}%
  \tdplotsetmaincoords{60}{-40}
\begin{tikzpicture}[tdplot_main_coords]

  \def\memristorWidth{0.5}
  \def\distanceBetweenMemristors{4*\memristorWidth}
  \def\numWordLines{5}
  \def\numBitLines{2}
  \tikzmath{\wordLineLength = \numBitLines*\distanceBetweenMemristors + \memristorWidth;}
  \tikzmath{\bitLineLength = \numWordLines*\distanceBetweenMemristors + \memristorWidth;}

  \foreach \j in {1,...,\numBitLines} {
    \boxColored{\j*\distanceBetweenMemristors}{2*\memristorWidth}{0}{\memristorWidth}{\memristorWidth}{\bitLineLength}{okabe-orange}{}{black};

    \draw[dashed, thick, okabe-orange] (\j*\distanceBetweenMemristors, 2*\memristorWidth, -0.5*\distanceBetweenMemristors) -- ++(\memristorWidth, 0, 0) -- ++(0, \memristorWidth, 0) -- ++(-\memristorWidth, 0, 0) -- cycle;
  }

  \foreach \i in {1,...,\numWordLines} {
    \foreach \j in {1,...,\numBitLines} {
      \boxColored{\j*\distanceBetweenMemristors}{\memristorWidth}{\i*\distanceBetweenMemristors}{\memristorWidth}{\memristorWidth}{\memristorWidth}{okabe-green}{}{black};
    }
  }

  \foreach \i in {1,...,\numWordLines} {
    \boxColored{0}{0}{\i*\distanceBetweenMemristors}{\wordLineLength}{\memristorWidth}{\memristorWidth}{okabe-blue}{}{black};

    \draw[dashed, thick, okabe-blue] (-0.5*\distanceBetweenMemristors, 0, \i*\distanceBetweenMemristors) -- ++(0, 0, \memristorWidth) -- ++(0, \memristorWidth, 0) -- ++(0, 0, -\memristorWidth) -- cycle;
  }

  \tikzmath{\zV = (\numWordLines+1)*\distanceBetweenMemristors/2 + 0.5*\memristorWidth;}
  \node[rotate around z=180, canvas is yz plane at x=0, font=\LARGE] at (0.5*\distanceBetweenMemristors, -2.5*\memristorWidth, \zV) {$\color{okabe-blue}\matr{V}$};

  \tikzmath{\xI = (\numBitLines+1)*\distanceBetweenMemristors/2;}
  \node[canvas is xy plane at z=0, font=\LARGE] at (\xI, 0, -0.5*\distanceBetweenMemristors) {$\color{okabe-orange}\matr{I^\top}$};

  \tikzmath{\xG = (\numBitLines+0.55)*\distanceBetweenMemristors;}
  \tikzmath{\zG = ((\numWordLines+1)/2 + 0.25)*\distanceBetweenMemristors;}
  \node[canvas is xz plane at y=0, font=\LARGE] (G) at (\xG, 0, \zG) {$\color{okabe-green}\matr{G}$};

  \node[right of=G, node distance=3cm, font=\large] (dot-product) {
    $
    \displaystyle
    \textcolor{okabe-orange}{I_{j}} = \sum_{i} \textcolor{okabe-blue}{V_{i}} \cdot \textcolor{okabe-green}{G_{i,j}}
    $
  };

  \node[below right of=dot-product, node distance=3cm, align=center] (multiplication) {
    \large Multiplication\\
    \small Ohm's law
  };
  \draw[->, thick] (multiplication) to [in=-90,out=90] ($(dot-product) + (0.99, 0, -0.3)$);

  \node[above right of=dot-product, node distance=3.5cm, align=center, xshift=-1.5cm] (addition) {
    \small Kirchhoff's current law \\
    \large Addition
  };
  \draw[->, thick] (addition) to [in=90,out=-90] ($(dot-product) + (-0.4, 0, 0.9)$);

  \node[below of=dot-product, node distance=4cm, font=\large] (matrix-vector) {
    $
    \displaystyle
    \textcolor{okabe-orange}{\matr{I^\top}} = \textcolor{okabe-blue}{\matr{V^\top}} \textcolor{okabe-green}{\matr{G}}
    $
  };

\end{tikzpicture}%

    \caption{
      Resistive crossbar array.
      The structure of the circuit together with Ohm's law and Kirchhoff's current law make it so that currents $\matr{I}$ are a product of a vector of voltages $\matr{V}$ and a matrix of conductances $\matr{G}$.
      Adapted from Ref.~\cite{mehonicEmergingNonvolatileMemories2023}.
    }
    \label{fig:crossbar-array}
\end{figure}

In-memory computing has been proposed as a solution to this problem.
For example, by exploiting Ohm's law, Kirchhoff's current law, and the structure of the circuit, resistive crossbar arrays are able to compute vector-matrix products, as shown in Figure~\ref{fig:crossbar-array}.
For fully connected networks, which are simply a series of blocks that perform this mathematical operation, this can lead to significant speedups.
Memristors are a promising candidate for crossbar arrays because their conductance can be easily tuned with voltage pulses, allowing to use them as proxies for the synaptic weights.
This can work especially well during inference, where high precision is less important.
This is similar to quantization, where models trained using high prevision can be compressed by using lower-precision representations, while minimizing the loss in accuracy.

But precision will always be a concern because resistive crossbars are inherently analog, thus perfect, digital-like computations are impossible.
Nonidealities, including stuck devices, $I$–$V$ nonlinearity, and line resistance, can all affect the accuracy in tasks like classification.
Different nonidealities vary in how difficult they are to model, predict, and mitigate, but techniques ranging from improved fabrication processes~\cite{HaAb2015,NaMo2016} to more sophisticated training algorithms~\cite{HeLi2019,ZhZh2020,JoLe2020} have been proposed to address these issues.

One of the issues rarely discussed in the context of analog machine learning accelerators is their security.
It is well established that statistical models like neural networks are vulnerable to adversarial attacks~\cite{szegedyIntriguingPropertiesNeural2014,biggio2017}, where---as an example of one type of attack---small perturbations to the input can cause misclassification.
If these devices are to be widely deployed in the real world, it is crucial to understand whether attackers could take advantage of memristive nonidealities to further \emph{exploit} vulnerabilities of these machine learning deployments or it would in fact make it \emph{more difficult} for them.
When it comes to adversarial robustness in memristive networks, it is of interest to 1) understand how the nonidealities affect the attacks, and 2) develop countermeasures specific to memristive networks.

In this work, we focus on a subset of memristive nonidealities related to stuck devices.
Many memristive devices operate on the principles of electroforming and switching, i.e., physical changes that cause changes in conductance.
As these changes happen at the level of individual atoms, they are prone to failure.
Additionally, stuck devices may approximate some noise injection techniques in traditional machine learning, which is relevant to the other focus of this work---nonideality-aware training.
We explore how this training technique, which is based on exposing nonidealities to the training algorithm, influences the network's ability to defend against attacks.

\section{Background and Design}

\subsection{Memristive neural networks}

Memristive devices have been extensively studied for inference acceleration.
Such devices encode weights in their conductance, which can be tuned with voltage pulses.
When arranged in an array structure, they can perform vector-matrix multiplication, a key operation in neural networks.
As a result, memristive devices are best suited for fully connected layers, though other implementations, such as convolutional layers, have been demonstrated in the past~\cite{YaWu2020}.
Inference (as opposed to training) is the more promising application, as the precision requirements are lower~\cite{Yu2018}.

After training on a conventional, digital platform, weights are typically transferred onto the memristive array.
Because conductances can only be positive, the weights are typically transferred to two sets of conductance arrays ($\matr{G}_+$ and $\matr{G}_-$) in the so-called differential-pair architecture.
For example, one may perform the following mapping from weight $w$ to conductances $G_+$ and $G_-$:

\begin{equation}\label{eq:weight-mapping}
  G_\pm = \frac{G_\mathrm{off} + G_\mathrm{on}}{2} \pm \frac{G_\mathrm{on} - G_\mathrm{off}}{2 \cdot \max|\matr{w}|} \cdot w
\end{equation}
where $G_\mathrm{off}$ and $G_\mathrm{on}$ are the lowest and highest achievable conductances, respectively, and $\max|\matr{w}|$ is the maximum absolute value of the weights~\cite{KiMa2021}.

Nonidealities may manifest in several ways, including in a nonlinear fashion.
For example, $I$-$V$ nonlinearities affect the input--output relationships, requiring to model devices individually, instead of assuming vector-matrix computation.
Similar assumptions hold for nonzero resistance along the wires connecting the devices, too, because they cause deviations from Ohmic behavior and thus make modelling more complicated.
On the other hand, device faults like stuck or unelectroformed devices can be modelled as deviations in $G$ (as long as Ohmic behavior holds true).

\subsection{Adversarial attacks}

In the context of machine learning, adversaries may aim to make a system behave in a certain way, disturb its operation, or compromise its privacy.

\textbf{Data poisoning} attacks, as an example, inject perturbed data during training to train a system to produce desired behavior later~\cite{10.1145/1128817.1128824,10.1145/2046684.2046692}.
\textbf{Evasion} attacks, on the other hand, aim to manipulate the input data to cause a misclassification in \emph{trained} models at inference time~\cite{szegedyIntriguingPropertiesNeural2014,biggio2017}.
In this work, we focus on evasion attacks.

\subsection{Attacking memristive networks}

Adversarial attacks on memristive networks are of great interest because of the unique set of constraints that these networks have.
Although much more efficient, they are subjected to physical nonidealities, which may have effects not only on network accuracy under normal conditions but also network robustness to attacks.
Also, the most popular way of utilizing these networks is by training them on a digital system and then mapping them onto the analog arrays of memristive devices.
However, this conductance programming stage often requires specialized hardware and may be unfeasible to perform once the networks have been deployed.
Thus, if a vulnerability is discovered, it may be impossible to deploy security patches, and the only feasible approach might be a recall.

Adversarial attacks on memristive networks have been considered in the past.
Both \citeauthor{bhattacharjeeRethinkingNonidealitiesMemristive2021}~\cite{bhattacharjeeRethinkingNonidealitiesMemristive2021} and \citeauthor{royIntrinsicRobustnessNVM2021}~\cite{royIntrinsicRobustnessNVM2021} have argued that memristive nonidealities offer intrinsic robustness to adversarial attacks with some of the explanations being gradient obfuscation~\cite{bhattacharjeeRethinkingNonidealitiesMemristive2021}.
Similarly, \citeauthor{paudelResiliencyAnalogMemristive2022}~\cite{paudelResiliencyAnalogMemristive2022} have argued that the introduction of additional nonideal analog components may lead to additional adversarial robustness.
\citeauthor{biADVNCSDversarial2024}~\cite{biADVNCSDversarial2024} explore how different training adjustments of memristive networks can produce greater adversarial robustness.

In this work, we expand on previous findings by exploring the robustness of memristive networks---with and without special training schemes and by also considering network designer's assumptions.
Specifically, we focus on the effect of stuck devices and nonideality-aware training when memristive networks are exposed to Fast Gradient Sign Method (FGSM)~\cite{goodfellowExplainingHarnessingAdversarial2015}.
This adversarial attack disturbs inputs by perturbing them in the direction of greatest loss increase:
\begin{equation}
    \matr{x}^\mathrm{adv} = \matr{x} + \varepsilon \cdot \mathrm{sign}(\nabla_{\matr{x}} J(\theta, \matr{x}, \matr{y}))
\end{equation}
where $J$ is a loss function.

For the simulations, we consider
\begin{itemize}
    \item network training assumptions (forward propagation function, which may take nonidealities into account)
    \item actual nonidealities encountered during inference
\end{itemize}

We use Fashion MNIST dataset~\cite{xiaoFashionMNISTNovelImage2017}, but all the qualitative results have been reproduced with MNIST dataset~\cite{LeCo2010} too and are available in the supplementary material.

\section{Results and Discussion}

\subsection{Effect of nonidealities}

\begin{figure}[h]
    \centering
    \includegraphics[width=\linewidth]{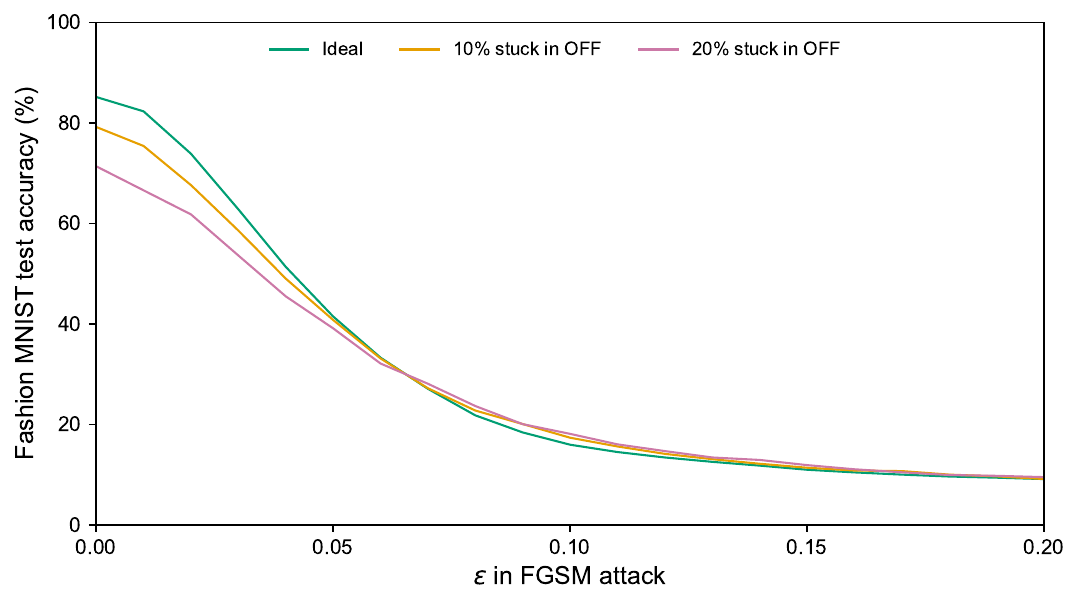}
    \caption{
        The effect of nonidealities in FGSM attack.
        A greater proportion of stuck devices results in lower accuracy under conditions of no attack.
        However, with high-$\varepsilon$ attack, the accuracy for all three scenarios becomes similar.
    }
    \label{fig:effect-of-nonidealities}
\end{figure}

Figure~\ref{fig:effect-of-nonidealities} shows how FGSM attack performs on ideal memristive networks, and those that have 10\% or 20\% random devices stuck in the OFF state.
Two qualitative features stand out: 1) the more stuck devices, the lower the starting accuracy (i.e., when $\varepsilon = 0$), and 2) as $\varepsilon$ increases, the accuracy for all three scenarios becomes similar.

The greater degree of nonidealities leading to lower accuracy in normal conditions is expected and has been reported in the past~\cite{MeJo2019}.
On the other hand, the effect of nonidealities diminishing as the attack magnitude is increased might be due to attacker's wrong assumptions.
In these attacks, the attacker assumes an ideal model in all cases---one that does not suffer from nonidealities and is effectively the same as digitally implemented machine learning model.
This is justified by the fact that 1) in each physical implementation, the distribution of stuck devices might be different, and 2) to know which devices are stuck, the attacker might need either physical access or more advanced attacks that extract this information.
The greater the degree of the nonidealities, the more the attacker's assumptions deviate from reality, and thus the attack becomes less effective as $\varepsilon$ is increased.

\subsection{Nonideality-aware training}\label{sec:nonideality-aware-training}

Nonideality-aware training is the idea of exposing nonidealities that are expected during inference to the training process.
In the case of, say, $20$\% stuck devices, through each training iteration, a different random sample of $20$\% devices would be set to OFF state during forward propagation.
Although the devices that are stuck during inference are different from any of the stuck devices in the training iterations, the network learns to adapt and becomes much more robust to this nonideality as a result.

\begin{figure}[h]
    \centering
    \includegraphics[width=\linewidth]{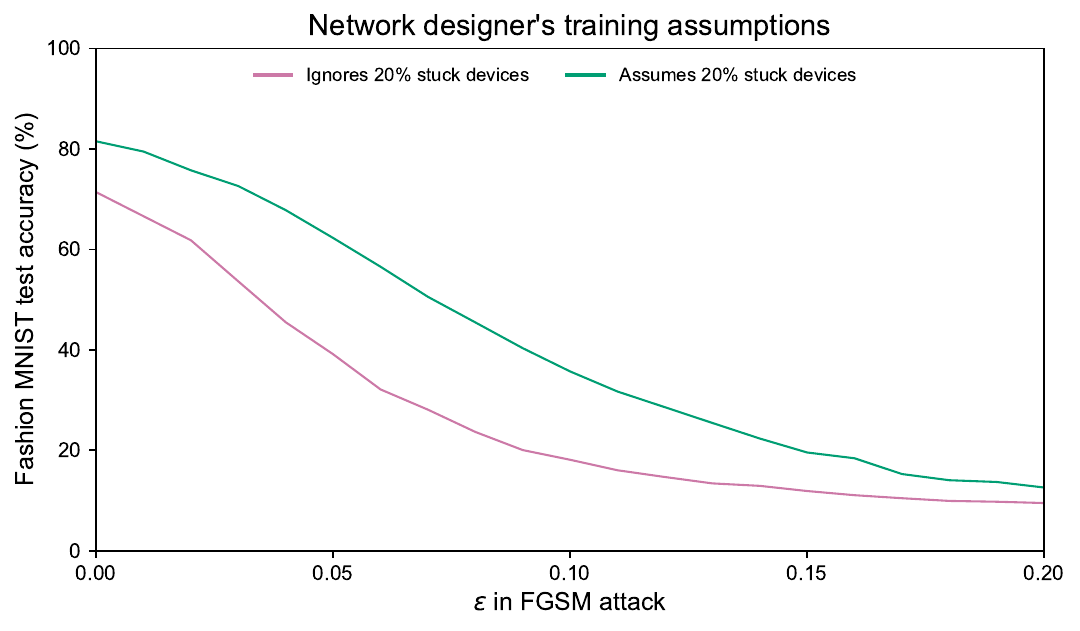}
    \caption{
        The effect of nonideality-aware on FGSM attack.
        Networks exposed to stuck devices have a much higher robustness to the attack.
    }
    \label{fig:aware-training}
\end{figure}

Figure~\ref{fig:aware-training} shows the robustness of nonideality-aware training to adversarial attacks, similar to the approach used by \citeauthor{biADVNCSDversarial2024}~\cite{biADVNCSDversarial2024}.
Unsurprisingly, already when no attack is present, training that assumes $20$\% stuck devices achieves higher accuracy during inference, because that is what it encounters.
But this higher accuracy advantage is maintained even as the attack magnitude is increased.

There are two possible explanations for why networks trained by exposing them to nonidealities might have higher robustness to attacks.
Firstly, it had been previously reported that networks with higher ``starting'' accuracy---the accuracy under ideal conditions---are more robust to changes during inference~\cite{MeJo2019}.
However, in that case, it could have been argued that it was due to the networks having a better representation of the dataset, while in this case, the networks were taught to better adapt to the \emph{nonidealities} during inference and not any particular aspect of the \emph{inputs}.
The second and the more likely explanation is that in the case of nonidealities like random stuck devices, exposing the network to this during training may be seen as noise injection.
In the past, noise injection has been shown to induce regularization and thus improve robustness to adversarial attacks~\cite{rakinParametricNoiseInjection2018}.

\subsection{Nonideality modelling during training}

Section~\ref{sec:nonideality-aware-training} assumed that the nonideality a designer models during training is the one encountered during inference.
In reality, there is uncertainty in how nonidealities may manifest in different samples of memristive devices.
Some might have lower yield (i.e., more stuck devices), others---higher.
Other nonidealities, say, $I$--$V$ nonlinearity, may too manifest to a different extent~\cite{JoWa2022}.

\begin{figure}[h]
    \centering
    \includegraphics[width=\linewidth]{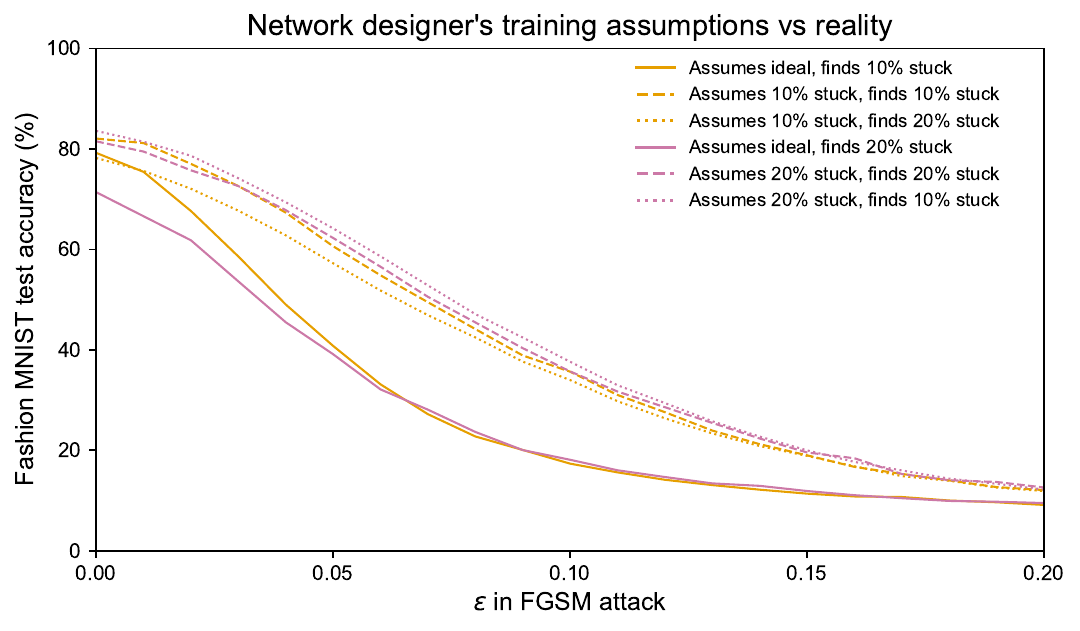}
    \caption{
        The effect of incorrect assumptions during network training.
        Not exposing the training algorithm to nonidealities has a negative effect on its robustness.
        However, even incorrect assumptions about \emph{some} level of nonidealities during inference produce higher robustness to attacks.
    }
    \label{fig:defender-assumptions}
\end{figure}

Figure~\ref{fig:defender-assumptions} considers incorrect assumptions when modelling nonidealities during training.
As seen in Section~\ref{sec:nonideality-aware-training}, correctly identifying inference-stage nonidealities during training leads not only to higher accuracy under no-attack conditions but also when an attack is happening.
Interestingly, as now demonstrated in Figure~\ref{fig:defender-assumptions}, making a wrong assumption about the \emph{magnitude} of an attack still produces significantly higher robustness to the attack.

An incorrect assumption about the \emph{level} of nonidealities is not only better than assuming no nonidealities but may in some cases even be better than guessing the level of nonidealities \emph{precisely}.
The robustness of nonideality-aware training with respect to modelling assumptions has been demonstrated in the past~\cite{JoWa2022}.
Because the \emph{nature} of these nonidealities ($10$\% stuck in OFF vs $20$\% stuck in OFF) is similar, it produces similar results during inference.
In the case of Figure~\ref{fig:defender-assumptions}, assuming $20$\% stuck devices while finding $10$\% stuck devices during inference produces even better results than assuming $10$\% stuck devices during training.
This can be explained by the fact that even with the correct assumptions about the \emph{level} of nonidealities, the exact distribution of stuck devices is not known until inference time.
Assuming higher proportion of stuck devices during training produces more robust networks and thus higher accuracy when stuck devices are encountered.

\section{Summary and conclusion}

Memristive networks are faster and more power efficient than their digital counterparts due to the elimination of the von~Neumann bottleneck.
However, the use of analog devices introduces a number of nonidealities that negatively affect the accuracy of these systems.
Techniques like nonideality-aware training may make the networks more robust to not only nonidealities but adversarial attacks too.

We have shown that when memristive neural networks suffer from stuck devices, nonideality-aware training makes them less vulnerable to FGSM attack even when nonideality modelling is imperfect.
As suggested by existing literature, we speculate that nonideality-aware training may act as noise injection and induce regularization, thus improving their robustness.
The fact that incorrect assumptions about nonidealities still produce advantageous results makes it an attractive option when design robust machine learning systems implemented using memristive devices.

\section*{Methods}

Fully connected networks used $784$ input neurons (plus bias), $32$ hidden neurons (plus bias), and $10$ output neurons.
In the memristive implementation, each weight was represented using a differential pair (two memristive devices).
The simulations used both Fashion MNIST (main text) and MNIST (supplementary material) datasets.
The networks were trained for $10$ epochs using cross-entropy loss.

Full implementation details can be found at\\
\texttt{\href{https://github.com/joksas/nonideality-aware-memristive-networks-under-attack}{github.com/joksas/nonideality-aware-memristive-networks-under-attack}}.

\printbibliography

\end{document}


\maketitle

\begin{figure}[h]
    \centering
    \includegraphics[width=\linewidth]{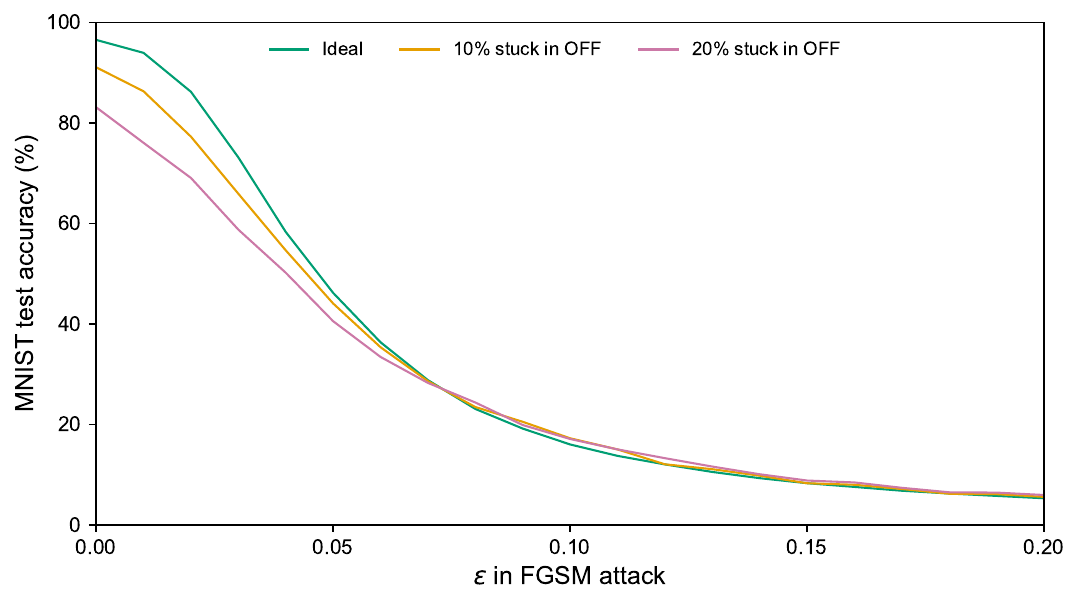}
    \caption{
        The equivalent of Figure~3 in the main text for the MNIST dataset.
    }
    \label{fig:effect-of-nonidealities}
\end{figure}

\begin{figure}[h]
    \centering
    \includegraphics[width=\linewidth]{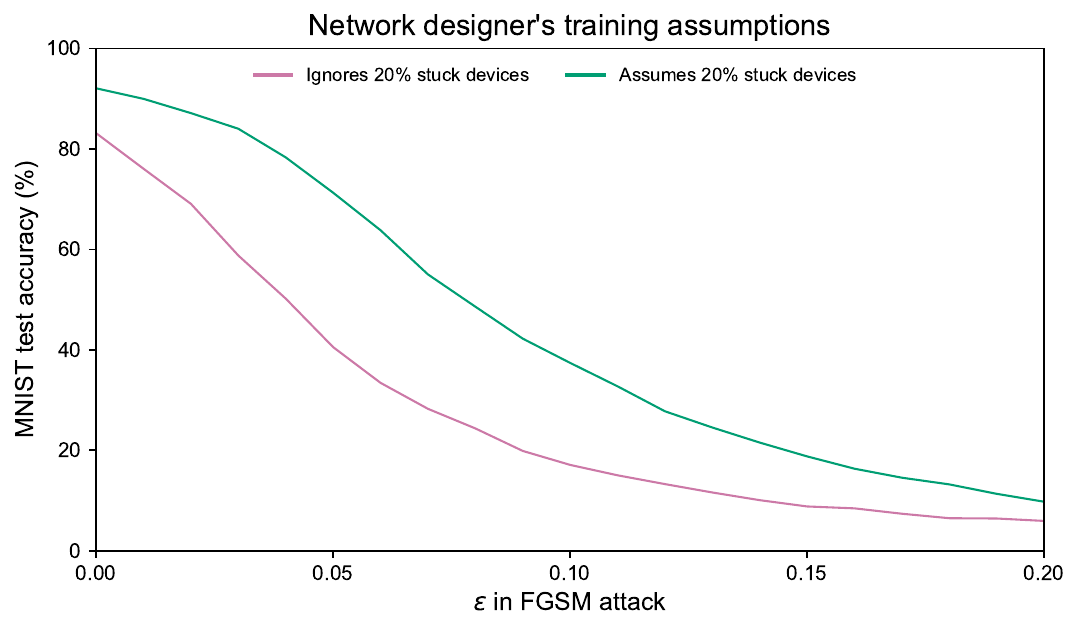}
    \caption{
        The equivalent of Figure~4 in the main text for the MNIST dataset.
    }
    \label{fig:aware-training}
\end{figure}

\begin{figure}[h]
    \centering
    \includegraphics[width=\linewidth]{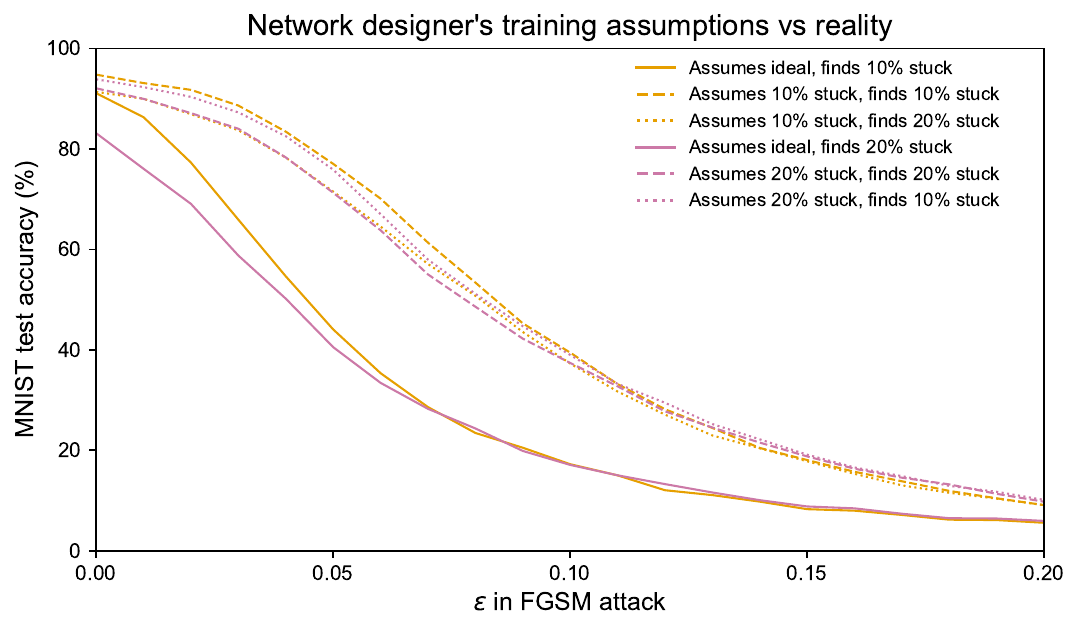}
    \caption{
        The equivalent of Figure~5 in the main text for the MNIST dataset.
    }
    \label{fig:defender-assumptions}
\end{figure}